\documentclass[twocolumn,showpacs,preprintnumbers,amsmath,amssymb,prl,superscriptaddress,longbibliography]{revtex4-1}
\usepackage{bbm}
\usepackage{mathrsfs}
\usepackage{graphicx}
\usepackage{dcolumn}
\usepackage{bm}
\usepackage{amsmath}
\usepackage{amsfonts}
\usepackage{color}
\usepackage[colorlinks=true,linkcolor=magenta,urlcolor=magenta,citecolor=cyan,anchorcolor=blue]{hyperref}
\usepackage{floatrow}

\begin{document}

\title{Giant anisotropic band flattening in twisted $\Gamma$ valley semiconductor bilayers}
\author{Huan Wang}
\thanks{Equal contribution.}
\affiliation{State Key Laboratory of Surface Physics and Department of Physics, Fudan University, Shanghai 200433, China}
\author{Zhaochen Liu}
\thanks{Equal contribution.}
\affiliation{State Key Laboratory of Surface Physics and Department of Physics, Fudan University, Shanghai 200433, China}
\author{Yadong Jiang}
\affiliation{State Key Laboratory of Surface Physics and Department of Physics, Fudan University, Shanghai 200433, China}
\author{Jing Wang}
\thanks{Corresponding author: wjingphys@fudan.edu.cn}
\affiliation{State Key Laboratory of Surface Physics and Department of Physics, Fudan University, Shanghai 200433, China}
\affiliation{Institute for Nanoelectronic Devices and Quantum Computing, Fudan University, Shanghai 200433, China}
\affiliation{Zhangjiang Fudan International Innovation Center, Fudan University, Shanghai 201210, China}

\begin{abstract}
We propose a general theory of anisotropic band flattening in moir\'e systems at {the $\Gamma$ valley. For a two-dimensional semiconductor with a rectangular unit cell of $C_{2z}$ or mirror symmetries, we find that a larger effective mass anisotropy $\eta=m_y/m_x$ of the valence or conduction bands in the monolayer will have a stronger tendency to be further enhanced in its twisted bilayer. This gives rise to strong anisotropic band flattening and correlated physics in one dimension effectively.} We predict twisted bilayer black phosphorus (tBBP) has \emph{giant} anisotropic flattened moir\'e bands ($\eta\sim10^4$) from \emph{ab} initio calculations and continuum model, where the low energy physics is described by the weakly coupled array of one-dimensional wires. We further calculate the phase diagram based on the sliding Luttinger liquid by including the screened Coulomb interactions in tBBP, and find a large parameter space may host  the non-Fermi liquid phase. We thus establish tBBP as a promising and experimentally accessible platform for exploring correlated physics in low dimensions.
\end{abstract}

\date{\today}

\maketitle

Moir\'e materials with flat electronic bands provide an ideal platform for exploring strongly correlated physics in two dimensions (2D)~\cite{andrei2020,balents2020,carr2020,kennes2021,andrei2021}. A paradigm example is twisted bilayer graphene at {the magic angle}~\cite{bistritzer2011}, which hosts flat bands and exhibits a variety of interacting phases including superconductors, correlated insulators and Chern insulators~\cite{cao2018b,cao2018a,yankowitz2019,sharpe2019,lu2019,serlin2020,nuckolls2020,xie2021}. Similar correlated phases have been observed in moir\'e systems of transition metal dichalcogenides~\cite{xu2020,foutty2023,cai2023,zeng2023} and multilayer graphene~\cite{chengr2019b,hao2021,park2021}. The correlated behaviors in moir\'e systems are associated with the quenched kinetic energy caused by moir\'e pattern, so strong electron interaction could dominate. 

An important correlated phase is the anisotropic non-Fermi liquid, which is the Luttinger liquid (LL) model generalized to higher dimension and could arise in 2D systems consisting of arrays of one-dimensional (1D) quantum wires~\cite{wen1990,emery2000,vishwanath2001,mukhopadhyay2001a,sondhi2001,sur2017,gao2021}. Experimentally realizing these coupled wire arrays perfectly is challenging. Most of the previous efforts have been made in the context of quasi-one-dimensional organic conductors~\cite{jerome1994,georges2000}. Recently, such anisotropic correlated phases may be observed in twisted bilayer WTe$_2$~\cite{wang2022}, which may host 1D flat bands~\cite{feng2023} and exhibit LL behavior in a 2D crystal. The anisotropic band flattening occurs when low-energy physics is located on the Brillouin zone edge but not at the zone corner~\cite{kariyado2019,kennes2020,koshino2022}, {and} is guaranteed by symmetry. However, realistic 2D materials fulfilling such requirements are extremely rare, while large classes of 2D semiconductors have valence band top or conduction band bottom occurring at {the $\Gamma$ valley, namely the Brillouin zone center}. This motivates us to study whether {the 1D flat band} could arise in twisted bilayer of the $\Gamma$ valley systems, and further identify realistic new platforms in which the LL phase and other exotic phases of matter may appear.

We develop the theory of anisotropic band flattening in a twisted bilayer of $\Gamma$ valley system. {For a 2D crystal with a rectangular unit cell of $C_{2z}$ or mirror symmetries, we find that a larger effective mass anisotropy of the valence or conduction bands in the monolayer will have a stronger tendency to be further enhanced in its twisted bilayer. Namely, the anisotropic band flattening occurs over most of the phase space if the monolayer has large effective mass anisotropy.} We propose twisted bilayer black phosphorus (tBBP) as a concrete example with giant anisotropic flattened moir\'e bands, where the low energy physics is described by the weakly coupled array of 1D wires from the real space charge density calculations. We further calculate the finite temperature phase diagram based on the screened Coulomb interactions, and propose tBBP as a promising and experimentally accessible platform for observing exotic non-Fermi liquid behavior in 2D.

{\color{cyan}\emph{Model.}}
We present band engineering by starting from the twisted bilayer with generic anisotropic band dispersions in each layer, where the low energy physics is at $\Gamma$. The general theory for the anisotropic band flattening presented here is generic for any semiconductors with anisotropic electronic dispersions {in the valence or conduction bands}. For simplicity, we first assume that each layer is a nonmagnetic rectangular lattice with a minimal $C_{2z}$ symmetry. We consider stacking of two identical layers taking $z$-axis as a normal direction. The top and bottom layer are rotated by {small angle} $+\varphi/2$ and $-\varphi/2$ around the $z$ axis, respectively. $\mathbf{a}_1$ and $\mathbf{a}_2$ are the Bravais unit vectors of monolayer rectangular lattice. Then the twisted moir\'e pattern has the periodicity of $\mathbf{L}_1$ and $\mathbf{L}_2$ as $\mathbf{L}_i=-\hat{\mathbf{z}}\times\mathbf{a}_i/(2\sin(\varphi/2))$. The generic effective model for electrons in such a moir\'e pattern is written as
\begin{equation}\label{model}
\mathcal{H}_{\text{eff}} = 
\begin{pmatrix} 
H_t(-i\boldsymbol\nabla) & \Lambda(\mathbf{r}) \\ 
\Lambda^\dagger(\mathbf{r}) & H_b(-i\boldsymbol\nabla) 
\end{pmatrix}.
\end{equation}
Here $H_{t/b}$ are the kinetic energy in top/bottom layer, which takes the general form as $H^\Gamma_{t/b}(\mathbf{k})=k_x^2/2m_x+k_y^2/2m_y$, with the momentum $k_{x/y}\in(-\pi/a_{1/2},\pi/a_{1/2}]$. The effective mass anisotropy is defined as $\eta\equiv m_y/m_x$. The moir\'e potential $\Lambda(\mathbf{r})$ is spatially periodic with the periodicity $\mathbf{L}_1$ and $\mathbf{L}_2$, which can be Fourier expanded by considering $C_{2z}$ and time reversal $\mathcal{T}$ symmetries of the $\Gamma$ valley as,
\begin{equation}\label{potential}
    \Lambda(\mathbf{r})=\Lambda_0+\sum_{n,i} \Lambda(\mathbf{g}_{n,i})\cos(\mathbf{g}_{n,i}\cdot\mathbf{r}),
\end{equation}
where $\mathbf{g}_{n,i}$ denote the moir\'e reciprocal lattice vectors to the $n$-th moir\'e Brillouin zone (mBZ) and $\Lambda(\mathbf{g}_{n,i})$ is real.
 
\begin{figure}[t]
\begin{center}
\includegraphics[width=3.4in,clip=true]{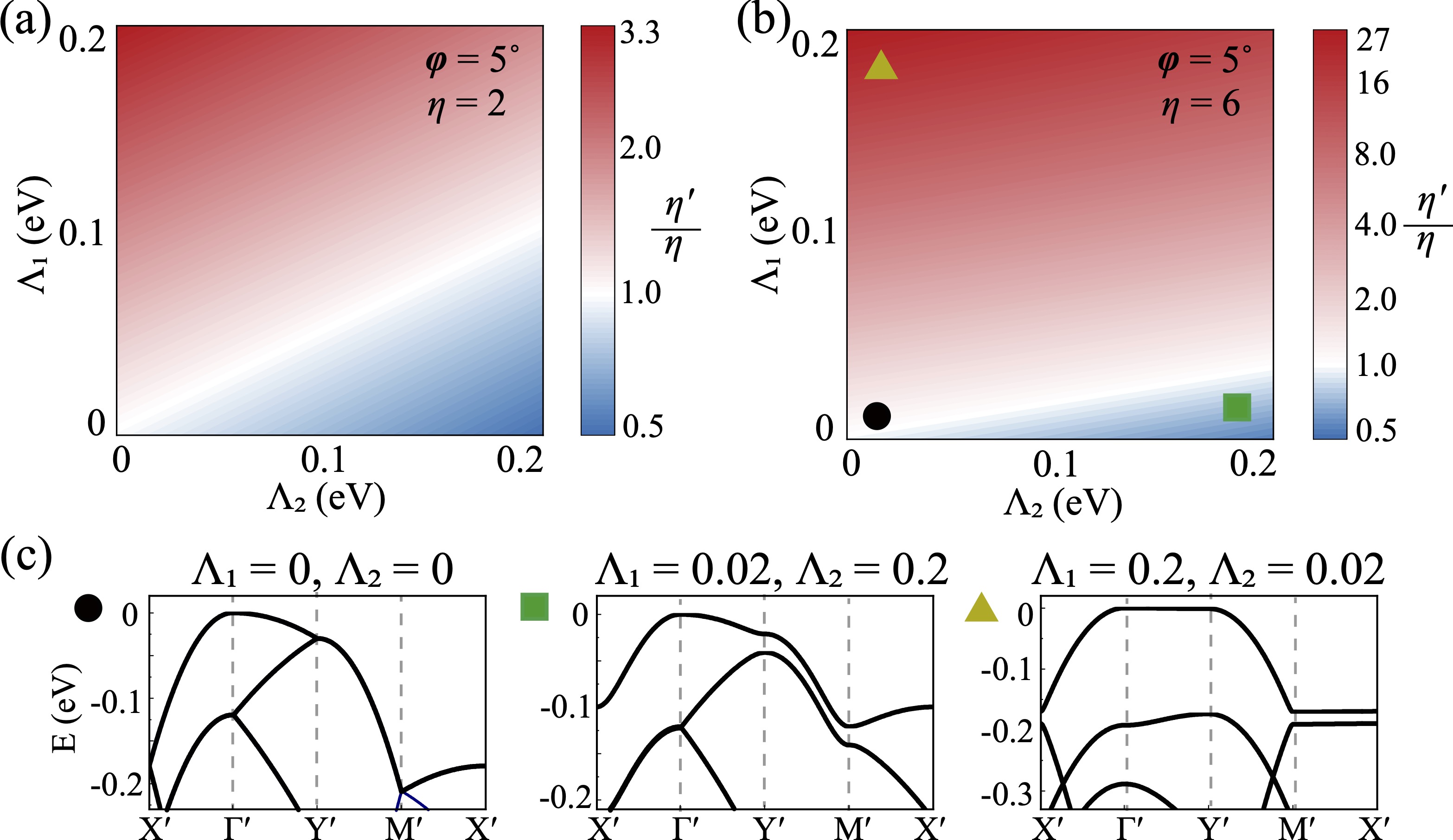}
\end{center}  
\caption{The evolution of effective mass anisotropy $\eta'$ of $\Gamma$ valley under moir\'e pattern. (a), (b) The ratio $\eta'/\eta$ vs $\Lambda_i$ for twist angle $\varphi=5^\circ$ and monolayer effective mass anisotropy $\eta=2$ and $\eta=6$, respectively. (c) Typical moir\'e band structures indicated by the symbols in (b). The anisotropic band flattening occurs in yellow triangle, and the topmost bands are two-fold degenerate from $\Lambda_0=0$. $a_1=4.0$~\AA, $a_2=3.9$~\AA.}
\label{fig1}
\end{figure}

{\color{cyan}\emph{Mass anisotropy under moir\'e pattern.}}
By diagonalizing Eq.~(\ref{model}), we obtain the moir\'e band structure. The low energy physics still occurs at $\Gamma$ and the effective Hamiltonian now becomes $H^{\text{moir\'e}}_{\Gamma}=k_x'^2/2m_x'+k_y'^2/2m_y'$, where $k'_{x/y}\in(-\pi/L_{2/1},\pi/L_{2/1}]$. Then we can see how effective mass anisotropy $\eta'\equiv m'_y/m'_x$ evolves under moir\'e potential. For simplicity, we only consider $\mathbf{g}_{n,i}$ up to the 1st mBZ, namely $n=1$, and $\Lambda(\mathbf{g}_{1,i})\equiv\Lambda_i$, we set $\Lambda_0=0$.

Fig.~\ref{fig1} shows the effective mass anisotropy ratio $\eta'/\eta$ as a function of $\Lambda_i$ at twist angle $\varphi=5^\circ$. In Fig.~\ref{fig1}(a), we set monolayer $\eta=2$, with $m_x=-0.1m_0$, $m_y=-0.2m_0$, and $m_0$ is the mass of free electron. In Fig.~\ref{fig1}(b), $\eta=6$, with $m_x=-0.1m_0$, $m_y=-0.6m_0$. In Fig.~\ref{fig1}(a) and~\ref{fig1}(b), $\eta'/\eta=1$ line is colored as white, while red and blue regions denotes the effective mass anisotropy is enhanced and suppressed under moir\'e potential, respectively. If the monolayer system is isotropic $\eta=1$, then $\eta'/\eta=1$ is along the diagonal line $\Lambda_1=\Lambda_2$ in phase space. Two conclusions can be drawn here. First, as the monolayer anisotropy $\eta$ gets larger, the area of anisotropy enhanced part ($\eta/\eta>1$) of the phase space {in the twisted bilayer} becomes bigger. This simply means the more anisotropy of the monolayer system is, the stronger tendency to get larger anisotropy in its twisted system. Second, as $\eta$ becomes larger, the effect of anisotropy enhancement gets better even when $\Lambda_i$ is the same. We can see the maximum of $(\eta'/\eta)_{\text{max}}=3.3$ in Fig.~\ref{fig1}(a), while in Fig.~\ref{fig1}(b) it becomes $(\eta'/\eta)_{\text{max}}=27$, even though $\eta$ of the monolayer is at the same order of magnitude. 

The above mass anisotropy behaviors can be understood from the band folding picture perturbatively. When $\Lambda(\mathbf{r})=0$, the folded band structure is shown in Fig.~\ref{fig1}(c) where the topmost {valence bands} are degenerate at Brillouin zone boundary X$'$, Y$'$ and M$'$. When $\Lambda(\mathbf{r})\neq0$, in the small angle limit, $H_{t/b}$ scales as $1/|\mathbf{L}_i|^2$ and becomes less important and $\Lambda(\mathbf{r})$ dominates. Then by treating $H_{t/b}$ as perturbations, we can diagonalize $\mathcal{H}_{\text{eff}}$ in which $\Lambda(\mathbf{r})$ is diagonalized,
\begin{equation}
\widetilde{\mathcal{H}}_{\text{eff}} = 
    \begin{pmatrix} 
    H_t(-i\boldsymbol\nabla)+\left|\Lambda(\mathbf{r})\right| & 0 \\ 
    0 & H_b(-i\boldsymbol\nabla)-\left|\Lambda(\mathbf{r})\right|
    \end{pmatrix}.
\end{equation}
We can see $\left|\Lambda(\mathbf{r})\right|$ plays a role of potential energy. $|\Lambda_1|$ and $|\Lambda_2|$ are the potential height along $\mathbf{a}_2$ and $\mathbf{a}_1$ axis, respectively, which open the gap at Y$'$ and X$'$. Intuitively, the potential tends to confine the electrons with large effective mass and small kinetic energy more effectively, while the electrons with small effective mass and large kinetic energy tends to move freely in such a potential. The moir\'e bandwidth with along $\Gamma'$-Y$'$ and $\Gamma'$-X$'$ are approximately $\pi^2/2|m_y|L_1^2-|\Lambda_1|/2$ and $\pi^2/2|m_x|L_2^2-|\Lambda_2|/2$, respectively. Thus the mass anisotropy ratio becomes
\begin{equation}\label{ratio}
\frac{\eta'}{\eta}\approx\frac{\pi^2-|m_x\Lambda_2|L_2^2}{\pi^2-|m_y\Lambda_1|L_1^2}\equiv\frac{1-\Lambda_2/2W_1}{1-\Lambda_1/2W_2},
\end{equation}
where $W_{1,2}\equiv\pi^2/2|m_{x,y}|L_{2,1}^2$ is the bandwidth along $\Gamma'$-X$'$,-Y$'$ of mBZ. {
It's not easy to see the relation between $\eta^\prime/\eta$ and $\eta$ directly. To further simplify Eq.~(\ref{ratio}), we set $L_1\sim L_2\sim L$ and $\Lambda_1\sim\Lambda_2\sim\lambda$, then
\begin{equation}\label{sim_ratio}
  \frac{\eta'}{\eta}\approx\frac{W-\lambda}{W-\lambda\eta},
\end{equation}
with $W\equiv\pi^2/m_xL^2$. Both of the numerator and denominator in Eq.~(\ref{sim_ratio}) are positive under the perturbative treatment, thus when $\eta>1$, $\eta^\prime/\eta>1$; and when $\eta\gg 1$, $\eta^\prime/\eta\gg 1$.} So the guiding principle to have a stronger anisotropic band flattening is to start with a larger anisotropy monolayer system.

\begin{figure}[t]
\begin{center}
\includegraphics[width=3.4in,clip=true]{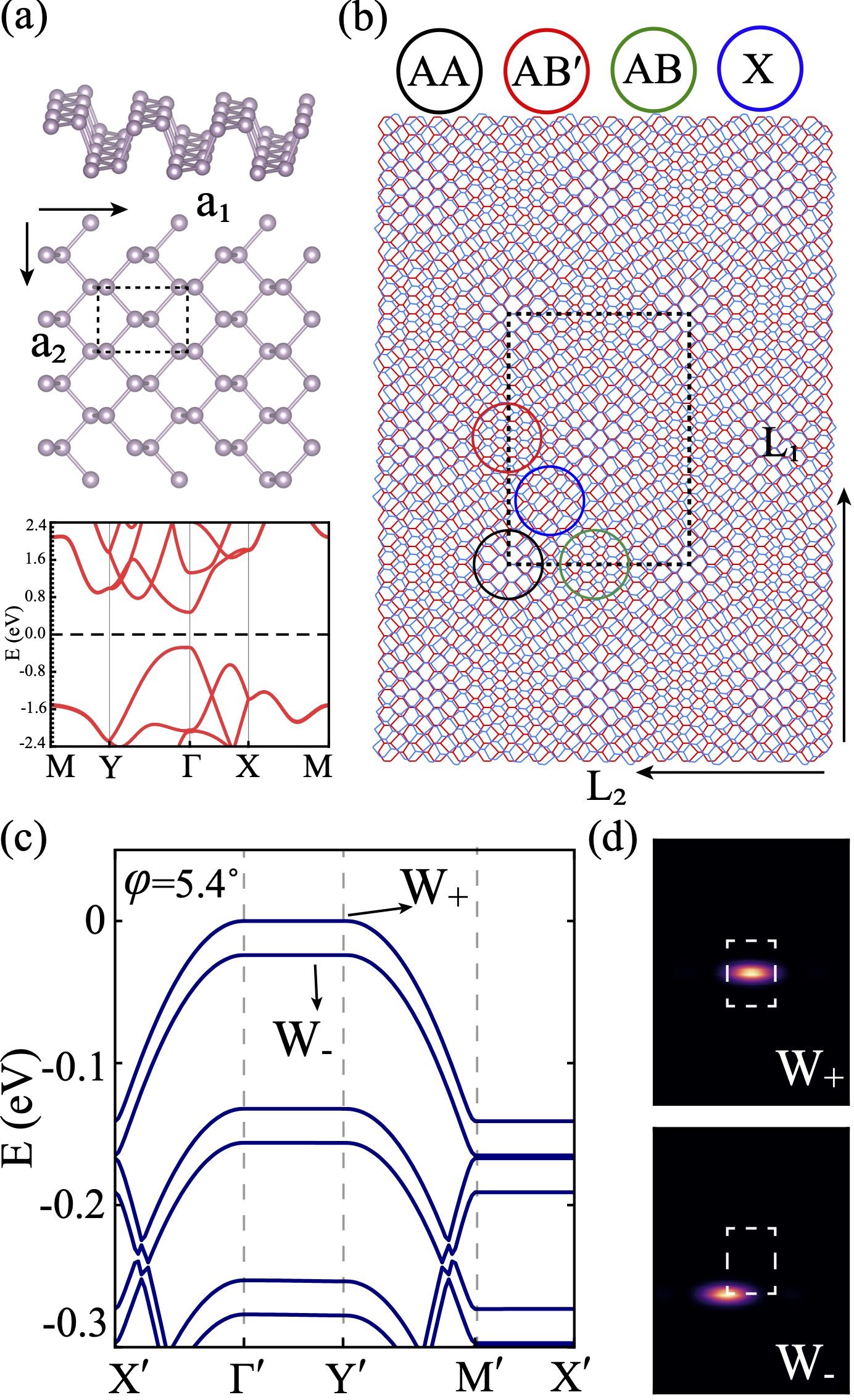}
\end{center} 
\caption{Small-angle tBBP moir\'e lattice and electronic properties. (a) Crystal structure and band structure of monolayer black phosphorus, the dashed rectangle indicates the unit cell. (b) Moir\'e superlattice formed on tBBP, where colored circles label several high-symmetry local stacking structures, with in-plane relative shift $\tau_{\text{AA}}=\mathbf{0}$, $\tau_{\text{AB'}}=\mathbf{a}_1/2$, $\tau_{\text{AB}}=\mathbf{a}_2/2$, $\tau_{\text{X}}=\mathbf{a}_1/4+\mathbf{a}_2/4$. (c) The moir\'e band structure from continuum model, and the Wannier functions of topmost two bands is shown in (d), where the white rectangle label the size of moir\'e unit cell.}
\label{fig2}
\end{figure}

{\color{cyan}\emph{Black phosphorus.}} To demonstrate the feasibility of our theory, we search for realistic materials. A paradigm example of 2D materials with strong anisotropic electronic dispersion is black phosphorus~\cite{lilikai2014,qiao2014}. The monolayer black phosphorus has a rectangular lattice with the space group $Pmna1^\prime$ (No.~53). The symmetry operations includes $\{C_{2z}|1/2,1/2\}$, $\{C_{2x}|1/2,1/2\}$, $M_y$ and $\mathcal{T}$. As shown in Fig.~\ref{fig2}(a), each primitive cell includes four phosphorus atoms. The low energy physics is around $\Gamma$ and is contributed from $p_z$ orbitals, where we find strong mass anisotropy $m_x=-0.2m_0$ and $m_y=-1.8m_0$ by fitting with \emph{ab} initio calculations in Fig.~\ref{fig1}(a), namely $\eta=9$.

Then we study the moir\'e band structure of tBBP. The moir\'e superlattice is shown in Fig.~\ref{fig2}(b). Quite different from moir\'e superlattice of tungsten in twisted WTe$_2$~\cite{wang2022}, there is no clear 1D stripes here. Since the translational symmetry is broken by the moir\'e pattern, the interlayer coupling $\Lambda(\mathbf{r})$ only respects $M_y$ and $\mathcal{T}$ symmetries. By expanding to the 1st mBZ, $\Lambda(\mathbf{r})$ can be expressed as 
\begin{eqnarray}\label{potential2}
\Lambda_{\text{BP}}(\mathbf{r}) &=& \Lambda_0+\sum_{i=1}^{4}\Lambda(\mathbf{g}_{1,i})e^{i\mathbf{g}_{1,i}\cdot\mathbf{r}},
\\
\Lambda(\mathbf{g}_{1,1}) &=& \Lambda^*(\mathbf{g}_{1,3})=\lambda_1+i\lambda_3,  \Lambda(\mathbf{g}_{1,2}) = \Lambda(\mathbf{g}_{1,4})=\lambda_2.\nonumber
\end{eqnarray} 
Here $\mathbf{g}_{1,1/2}$ is the moir\'e reciprocal lattice vectors to the 1st mBZ, $\mathbf{g}_{1,3/4}=-\mathbf{g}_{1,1/2}$. $\Lambda_0$ and $\lambda_j$ ($j=1,2,3$) are real parameters, which can be obtained by fitting the band structures from \emph{ab} initio calculations with untwisted but shifted configurations such as AA, AB$^\prime$, AB and X shown in Fig.~\ref{fig2}(b)~\cite{jung2014,supple}. We find $\Lambda_0=12$ meV, $\lambda_1=137$ meV, $\lambda_2=13$ meV, and $\lambda_3=-10$ meV.

The band structure of tBBP with $\varphi=5.4^\circ$ is shown in Fig.~\ref{fig2}(c), where two topmost valence bands have a \emph{giant} anisotropic band flattening of $\eta'\approx2\times10^4$ as shown in Fig.~\ref{fig4}(c). These two bands are only dispersive along $\Gamma'$-Y$'$ (or X$'$-M$'$) direction and dispersionless along the perpendicular $\Gamma'$-X$'$ (or Y$'$-M$'$) direction. Similar band structure from \emph{ab} initio calculations is obtained in Supplementary Materials~\cite{supple,kang2017,kang2020}. In Fig.~\ref{fig2}(c), the topmost two valence bands are from the trapping at two local maximum points of moir\'e potential $|\Lambda_{\text{BP}}(\mathbf{r})|$, which are located at the corner and middle of unit cell, respectively. The potentials at these two points are different due to finite $\Lambda_0$, thus leads to splitting of the two topmost bands compared to degenerate bands with $\Lambda_0=0$ in Fig.~\ref{fig1}(c). We further construct the maximally localized Wannier functions of these two bands~\cite{marzari1997}. As shown Fig.~\ref{fig2}(d), we find Wannier state of top band W$_+$ is localized at the center of moir\'e unit cell, and represents a layer bonding state for $\Lambda_{\text{BP}}(\mathbf{r})$ is positive. While the Wannier state of the second topmost band W$_-$ is localized at the corner of moir\'e unit cell, and is a layer anit-bonding state for negative $\Lambda_{\text{BP}}(\mathbf{r})$. Both of the Wannier states are highly anisotropic in real space.

\begin{figure}[t]
\begin{center}
\includegraphics[width=3.4in,clip=true]{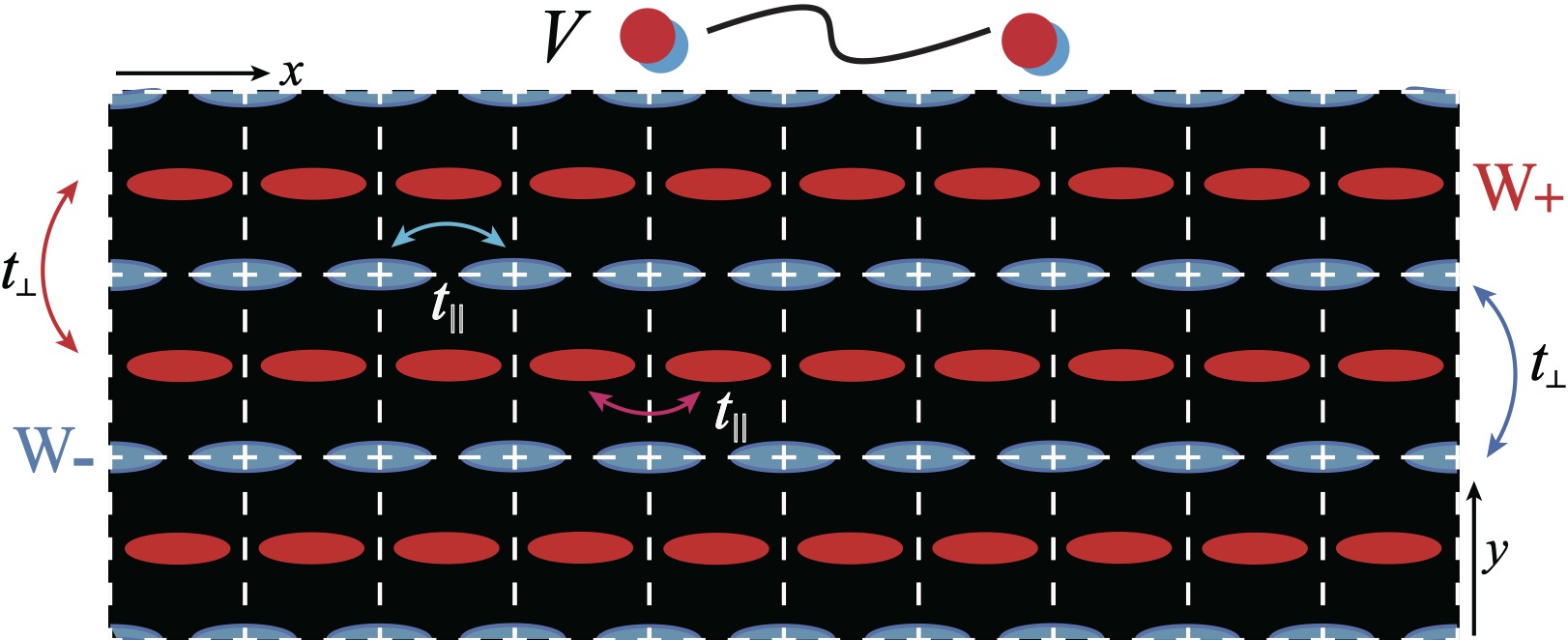}
\end{center} 
\caption{Schematics of array of parallel weakly coupled 1D wires. White dashed rectangle denotes the moir\'e unit cell, red ellipse is the layer bonding state W$_+$ at the center of unit cell, blue ellipse is the layer anti-bonding state W$_-$ at the corner of unit cell. $t_{\perp}$ and $t_{\parallel}$ are the interwire and intrawire hopping, respectively. $V$ is the long range Coulomb interaction.}
\label{fig3}
\end{figure}

\begin{figure}[t]  
\begin{center}
\includegraphics[width=3.4in,clip=true]{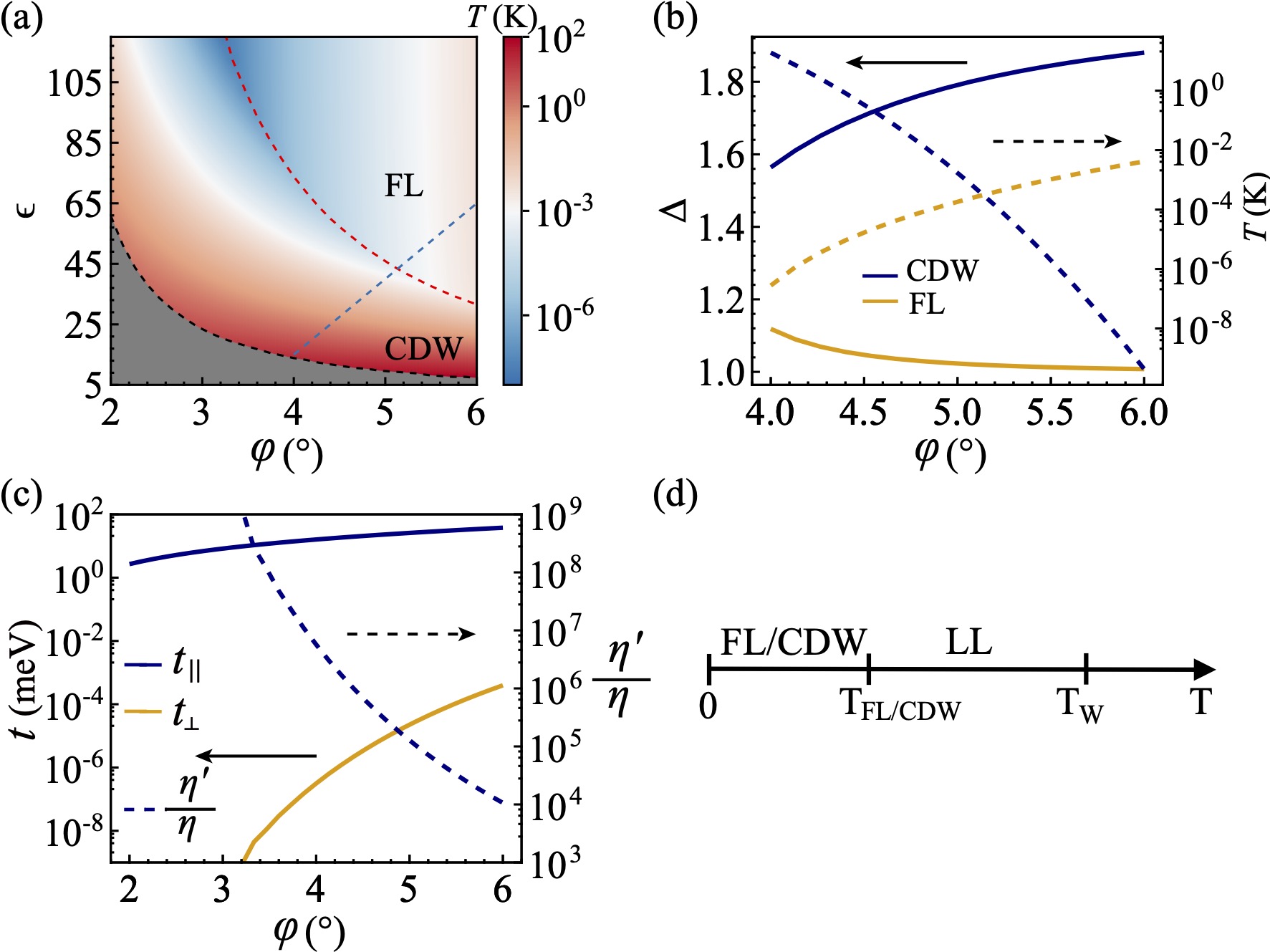}
\end{center} 
\caption{(a) Finite temperature phase diagram vs dielectric constant $\epsilon$ and twist angle $\varphi$ in tBBP. Color shows the transition temperature of the corresponding phase. For the shaded region, the Hubbard interaction is larger than the bandwidth, which is inaccessible by bosonization approach. (b) The scaling dimension (solid lines) and transition temperature (dashed lines) of CDW and FL states along the blue dashed line in (a). The single particle tunneling is most relevant. (c) $t_{\parallel}$, $t_{\perp}$ and $\eta'/\eta$ vs $\varphi$ in tBBP. (d) Schematic of dimensional crossover from sliding LL to CDW/FL.}
\label{fig4}
\end{figure}

{\color{cyan}\emph{Weakly coupled 1D wires array.}}
With giant anisotropic flatten moir\'e bands, tBBP now is descried by the array of weakly coupled 1D wires illustrated in Fig.~\ref{fig3}, which can expand the LL physics to 2D. Here we study the effect of Coulomb interaction in this system. The lattice Hamiltonian is constructed by projecting the single particle model and Coulomb interaction onto the above two maximally localized Wannier states. Since these Wannier states are well localized and separated from each other, we only keep the onsite Hubbard term and long-ranged density-density interaction, and other interactions such as exchange, paired-hopping and correlated hopping have been neglected. The bosonized form of the 1D fermion operator is $\psi_{s,r,j}(x)=(\zeta_{r,s,j}/\sqrt{2\pi \varepsilon})e^{irk_F x}e^{-i(r\phi_{s,j}-\theta_{s,j})}$, where $\varepsilon$ is related to intrachain cutoff with dimension of length, $r=\pm$ stands for the right and left moving electrons, $s= \uparrow,\downarrow$ is the spin index, $j$ denotes the wire number, $\zeta_{r,s,j}$ is the Kelvin factor, $\phi_{s,j}$ is the density variable and $\theta_{s,j}$ is the conjugate phase variable~\cite{von1998,giamarchi2003}. The boson operators for charge and spin excitations are given by $\phi_{\rho,j}=(\phi_{\uparrow,j}+\phi_{\downarrow,j})/\sqrt{2}$ and $\phi_{\sigma,j}=(\phi_{\uparrow,j}-\phi_{\downarrow,j})/\sqrt{2}$, respectively, and a similar relation for the dual $\theta$ field. For the array of 1D wires, we can write down an effective Hamiltonian as
\begin{eqnarray}
\mathcal{H}_{\text{1d}} &=& \frac{1}{2\pi\Omega a_y}\sum_{q}\sum_{\beta=\rho,\sigma}u_{\beta}q^2_x\Big[K_{\beta}(q)\theta_{\beta,q}\theta_{\beta,-q}
\\
&+&\frac{1}{K_{\beta}(q)}\phi_{\beta,q}\phi_{\beta,-q}\Big]+\sum_{j}\int dx\frac{2Ua_x}{(2\pi\varepsilon)^2}\cos{\sqrt{8}\phi_{j,\sigma}},
\nonumber
\label{1d_array}
\end{eqnarray} 
where $a_x$ and $a_y$ are intrachain and interchain lattice constant, respectively. $\Omega$ is the system area, $U$ is Hubbard interaction strength. We assume the non-commensurate filling, thus the umklapp process in the charge sector of Hubbard term cannot exist. Also, we only keep the forward scattering in the long range density-density interaction~\cite{giamarchi2003}. The Luttinger parameter and velocity are $u_{\rho}K_{\rho}=v_F,u_{\rho}/K_{\rho}=v_F+Ua_x/\pi+2V(q)/\pi a_y$ and $u_{\sigma}K_{\sigma}=v_F,u_{\sigma}/K_{\sigma}=v_F-Ua_x/\pi$. $V(q)$ is the Fourier transformed Coulomb interaction. Here we consider the single gate screened Coulomb interaction with $V(q)=(1-e^{-2d_sq})e^2/2\epsilon\epsilon_0q$, where $d_s$ is the distance between the sample and gate, and we set $d_s=50$~nm, $\epsilon$ is dielectric constant of the substrate. The spin sector with the bare Luttinger parameter in the attractive interaction regime (i.e. $K_{\sigma}>1$) flows to the gapless phase with $K_{\sigma}=1$ under renormalization group (RG), which is imposed by spin SU(2) symmetry. Thus the last term in $\mathcal{H}_{\text{1d}}$ vanishes~\cite{giamarchi2003}. 

Now we can see $\mathcal{H}_{\text{1d}}$ with the remaining first term is just the generic form of sliding LL. The instabilities of this model under various interaction couplings have been studied perviously~\cite{emery2000,mukhopadhyay2001a,vishwanath2001,sondhi2001,sur2017}. The most relevant interactions are interchain electron tunneling (ET), charge-density wave (CDW) and superconducting couplings (SC), which are expressed as
\begin{eqnarray}
&&\mathcal{H}_{{\text{ET}},n} \propto \sum_{s,r,j}\int dx\psi^\dagger_{s,r,j}\psi_{s,r,j+n}+\text{H.c.}
\\
&&\propto \sum_{j}\int dx \sum_{s=\uparrow,\downarrow}\cos{(\phi_{s,j}-\phi_{s,j+n})}\cos{(\theta_{s,j}-\theta_{s,j+n})},
\nonumber
\\ 
&&\mathcal{H}_{{\text{SC}},n} \propto \sum_{j}\int dx (\psi^\dagger_{R,\uparrow,j}\psi^\dagger_{L,\downarrow,j}+\psi^\dagger_{R,\downarrow,j}\psi^\dagger_{L,\uparrow,j})\times
\nonumber
\\
&&\quad\quad\quad (\psi_{R,\uparrow,j+n}\psi_{L,\downarrow,j+n}+\psi_{R,\downarrow,j+n}\psi_{L,\uparrow,j+n})+\text{H.c.}
\nonumber
\\
&&\propto\sum_{j}\int dx \cos{\sqrt{2}(\theta_{\rho,j}-\theta_{\rho,j+n})}\cos{\sqrt{2}\phi^{\sigma}_{j}}\cos{\sqrt{2}\phi^{\sigma}_{j+n}},
\nonumber
\\
&&\mathcal{H}_{{\text{CDW}},n} \propto \sum_{j}\int dx (\psi^\dagger_{R,\uparrow,j}\psi_{L,\uparrow,j}+\psi^\dagger_{R,\downarrow,j}\psi_{L,\downarrow,j})\times
\nonumber
\\
&&\quad\quad\quad\quad (\psi^\dagger_{L,\uparrow,j+n}\psi_{R,\uparrow,j+n}+\psi^\dagger_{L,\downarrow,j+n}\psi_{R,\downarrow,j+n})+\text{H.c.}
\nonumber
\\
&&\propto\sum_{j}\int dx \cos{\sqrt{2}(\phi_{\rho,j}-\phi_{\rho,j+n})}\cos{\sqrt{2}\phi^{\sigma}_{j}}\cos{\sqrt{2}\phi^{\sigma}_{j+n}}.
\nonumber
\end{eqnarray}
Here $\mathcal{H}_{{\alpha},n}$ ($\alpha=$ ET, SC, CDW) describe the coupling between chains separated by a distance of $na_y$. After projecting the Coulomb interaction $V(q)$ along the chain by $V(q_x=0,q_y)$, we can obtain the scaling dimension $\Delta_{\alpha,n}$ for the coupling strength $J_{\alpha,n}$ of $H_{\alpha,n}$ as
\begin{eqnarray}
\Delta_{\text{CDW},n} &=& \int_{-\pi}^\pi \frac{dq_y}{2\pi}(1-\cos(nq_y))K_{\rho}(q_y)+1,
\nonumber
\\
\Delta_{\text{SC},n} &=& \int_{-\pi}^\pi \frac{dq_y}{2\pi}(1-\cos(nq_y))K^{-1}_{\rho}(q_y)+1,
\nonumber
\\
\Delta_{\text{ET},n} &=& \frac{1}{4}(\Delta_{\text{SC},n}+\Delta_{\text{CDW},n}).
\end{eqnarray}

{\color{cyan}\emph{Phase diagram and sliding LL.}}
The screened Coulomb interaction decays rapidly in real space when the distance exceeds the screening length. Thus it is legitimate to consider only the nearest neighbour ($n=1$) coupling for CDW and SC couplings, while the leading order for electron tunneling is $n=2$, because the tunneling between W$_+$ and W$_-$ is exactly zero. The subscript $n$ is now omitted. By varying the dielectric constant $\epsilon$ and twisting angle $\varphi$, we find the electron tunneling is the most relevant ($\Delta_{\text{ET}} \approx 1.1$) and the CDW coupling is subdominant ($\Delta_{\text{CDW}} \approx 1.7$), while the superconductivity is always irrelevant ($\Delta_{\text{SC}}>2$) [see Fig.~\ref{fig4}(b)]. Therefore, we expect in the array of 1D wires a crossover from non-Fermi liquid (namely sliding LL) behavior to Fermi liquid or CDW as the temperature is lowered as shown in Fig.~\ref{fig4}(d).

The finite temperature phase diagram in tBBP where the crossover takes place is calculated in Fig.~\ref{fig4}(a). The interactions renormalize the coupling strength, and upon rescaling the RG equation for $J_\alpha$ at tree level, we obtain $dJ_{\alpha}/dl=J_{\alpha}(2-\Delta_{\alpha})$. {$J_{\text{ET}}=t_{\perp}$ and $J_{\text{CDW}}\approx e^2/4\pi\epsilon\epsilon_0a_y$}. We determine the crossover scale by the energy at which the renormalized coupling strength is of order one $J_\alpha/W\sim1$ where $W=4t_{\parallel}$ is the order of intrachain bandwidth. Thus, $T_{\alpha}\sim W(J_{\alpha}/W)^{1/(2-\Delta_{\alpha})}$~\cite{georges2000,giamarchi2003}. Since $\Delta_\alpha$ is always larger than one for interacting system, we see that the temperature scale at which the crossover takes place is always smaller than the non-interacting case $T_\alpha\sim J_\alpha$. As shown in Fig.~\ref{fig4}(a), although interchain single particle tunneling is the most relevant, there is still quite a region in the phase diagram where the transition temperature of CDW phase is larger than that of Fermi liquid (FL), caused by interchain electron tunneling. Moreover, the energy scale of the intrachain bandwidth $T_{\text{W}}$ in Fig.~\ref{fig4}(c) is ranging from $90$~K ($\varphi=2^\circ$) to $1700$~K ($\varphi=6^\circ$). Therefore, we expect the LL behaviour can be observed over a {wide} temperature range {($T_{\text{CDW/FL}}<T<T_{\text{W}}$)} in tBBP. In shaded region of Fig.~\ref{fig4}(a), the onsite Hubbard $U$ is much larger than the bandwidth (thus $K_{\sigma}^2<0$), which invalidates the bosonization approach and the system is considered as the Hubbard model, and this is beyond the current scope of the study. 

{\color{cyan}\emph{Summary.}} In summary, we propose the guiding principle for a stronger anisotropic band flattening in the $\Gamma$ valley moir\'e system is to start with a larger anisotropy monolayer system, such as but not limited to the rectangular lattice with $C_{2z}$ or mirror symmetries. We predict tBBP as a paradigm example with giant anisotropic flattened moir\'e bands, which provides a highly tunable platform for studying the weakly coupled 2D array of 1D electronic structures. The interchain hopping with several order of magnitude smaller compared to intrachain hopping in tBBP gives rise to a {wide} temperature range to observe the LL behavior in 2D. The twisted multilayer black phosphorus is also expected to host anisotropic flatten 1D bands. The rich choice of $\Gamma$ valley 2D semiconductors (for example, black arsenic~\cite{sheng2021}) provides great opportunities for studying many novel correlated and topological quantum phases~\cite{emery2000,vishwanath2001,mukhopadhyay2001a,sondhi2001,mukhopadhyay2001b,kane2002,teo2014,neupert2014,tam2021}.

\begin{acknowledgments}
We acknowledge helpful discussions with Biao Lian and Yuanbo Zhang. This work is supported by the National Key Research Program of China under Grant No.~2019YFA0308404, the Natural Science Foundation of China through Grant No.~12174066, the Innovation Program for Quantum Science and Technology through Grant No.~2021ZD0302600, Science and Technology Commission of Shanghai Municipality under Grant No.~20JC1415900, Shanghai Municipal Science and Technology Major Project under Grant No.~2019SHZDZX01.
\end{acknowledgments}

\end{document}